\documentclass[sigconf]{acmart}

\AtBeginDocument{%
  }

\copyrightyear{2025}
\acmYear{2025}
\setcopyright{cc}
\setcctype{by}
\acmConference[CHI '25]{CHI Conference on Human Factors in Computing Systems}{April 26-May 1, 2025}{Yokohama, Japan}
\acmBooktitle{CHI Conference on Human Factors in Computing Systems (CHI '25), April 26-May 1, 2025, Yokohama, Japan}\acmDOI{10.1145/3706598.3713729}
\acmISBN{979-8-4007-1394-1/25/04}

\begin{document}

\title{“A Bridge to Nowhere”: A Healthcare Case Study for Non-Reformist Design}


\author{Linda Huber}
\affiliation{%
 \institution{University of Michigan}
 \city{Ann Arbor}
 \state{Michigan}
 \country{United States}}

\begin{abstract}
In the face of intensified datafication and automation in public-sector industries, frameworks like design justice and the feminist practice of refusal provide help to identify and mitigate structural harm and challenge inequities reproduced in digitized infrastructures. This paper applies those frameworks to emerging efforts across the U.S. healthcare industry to automate prior authorization - a process whereby insurance companies determine whether a treatment or service is “medically necessary” before agreeing to cover it. Federal regulatory interventions turn to datafication and automation to reduce the harms of this widely unpopular process shown to delay vital treatments and create immense administrative burden for healthcare providers and patients. This paper explores emerging prior authorization reforms as a case study, applying the frameworks of design justice and refusal to highlight the inherent conservatism of interventions oriented towards improving the user experience of extractive systems. I further explore how the abolitionist framework of non-reformist reform helps to clarify alternative interventions that would mitigate the harms of prior authorization in ways that do not reproduce or extend the power of insurance companies. I propose a set of four tenets for non-reformist design to mitigate structural harms and advance design justice in a broad set of domains. 
\end{abstract}



\received{12 September 2024}

\maketitle
\vspace{5mm} 

\section{Introduction}
As automated decision-making and data-driven business models become increasingly prevalent within public sectors like healthcare, transportation, housing and agriculture, frameworks which help to identify and mitigate the structural harms and inequities of technological systems are increasingly crucial. Within HCI scholarship, a growing body of work points to the importance not only of “data ethics,” or mitigating bias and harm within specific datasets or algorithms, but of “data justice,” which calls us to challenge the power inequities and structural harms within broader sociotechnical systems \cite{Costanza-Chock2020-bz, DIgnazio2023-vu}. Related scholarship explores how the feminist practice of “refusal” can inform designers and technologists, including via the generative, creative aspects of refusal \cite{Garcia2022-iz}. 

Through a concrete case study of an emerging process of automation in the U.S. healthcare industry, this paper explores what it looks like to move towards more systemic or structural forms of data justice - including the very real difficulty of balancing immediate harm reduction with long-term systemic change. This paper specifically explores the framework of “non-reformist reform,” initially outlined by social theorist Andre Gorz  \cite{gorz1968strategy} and developed and advanced primarily by prison abolitionists. The framework of non-reformist reform helps to describe the creative challenge of identifying small, actionable steps towards systemic change \cite{Akbar2022-ko, Gilmore2007-cv} - or at least, to identify ways to reduce immediate social harms without perpetuating inequities and injustice.

As a case study in non-reformist reforms, this paper critically examines emerging efforts to automate “prior authorizations,” a process whereby insurance companies determine whether a treatment or service is “medically necessary” before agreeing to cover it. Prior authorizations are intended to mitigate rising healthcare costs - but mounting evidence of the harm caused to both patients and providers has made it the subject of reformist interventions by state and federal regulators. This paper argues that the automation of prior authorization may have a short-term harm reduction impact - but nonetheless extends and reproduces larger systemic harms and inequities. Specifically, I argue that the automation of prior authorization helps to extend insurance companies’ power to control care in ways that are fundamentally extractive. I apply the framework of non-reformist reforms to help imagine alternative, incremental changes that would help to reduce the immediate harms of prior authorization without extending the power of insurance companies over care. 

More broadly, this paper explores the ways that technological solutions oriented towards greater efficiency and reduction of user burden tend to be fundamentally reformist in nature. I demonstrate the utility of non-reformist reforms as a framework for designers and technologists looking for ways to advance design justice, particularly in the face of intensified datafication and automation in public sectors and industries. I point towards theories and practices of abolitionist activists and scholars as a powerful resource for thinking about creative ways to seed alternative futures within our work today.

This paper proceeds as follows. I first review frameworks proposed in HCI and critical data studies literature that highlight the limits of proceduralist approaches to “ethics,” and point towards the necessity of both technical- and non-technical practices that grapple with the structural violence of overall systems. I then provide a brief overview of the 18+ months of qualitative fieldwork on the datafication of the U.S. healthcare industry from which the current case study is drawn. I then move into the case study itself, providing a detailed overview of the intent and impact of prior authorization as a mechanism of rationing care - and the intent and likely impact of technological reforms to this system. This overview lays the groundwork for my evaluation of prior authorization as part of a fundamentally unjust, extractive system - and its proposed automation as a kind of reformist-reform. I then provide four proposed tenets of non-reformist design to help think through how else the immediate harms of prior authorization might be addressed without reproducing the injustice of the current system for adjudicating “medically necessary care.” I conclude by exploring how non-reformist reforms can help inform the work of design, particularly given the inherent conservatism of fixes rooted in improved “user experience” and technological efficiency.

\section{From Data Ethics to Data Justice: Frameworks for Reformist versus Systemic Change through Design}

Within critical data studies and feminist HCI scholarship there have been multiple frameworks introduced to distinguish between design practices that are largely conservative - in the sense that they are averse to change or preserve traditional institutions - and those that more fundamentally challenge an overall system. D’Ignacio and Klein, for instance, contrast \textit{data ethics} as a framework that “secures power” by “locat[ing] the source of the problem in individuals or technical systems” with \textit{data justice} as a concept that challenges power differentials and “work[s] towards dismantling them” \cite{DIgnazio2023-vu}.   Green and Viljoen \cite{green2020algorithmic} point to a similar distinction in Dunne and Raby's foundational framework of "critical design": 
\begin{quote}
    Design can be described as falling into two very broad categories: affirmative design and critical design. The former reinforces how things are now, it conforms to cultural, social, technical and economic expectation. Most design falls into this category. The latter rejects how things are now as being the only possibility, it provides a critique of the prevailing situation through designs that embody alternative social, cultural, technical or economic values. (p.58) \cite{Dunne2001-rq} 
\end{quote} 
In a poignant caricature of this distinction between the type of framework that 'reinforces how things are now' and one that 'provides a critique of the prevailing system," Keyes, Hutson and Durbin outline a proposal for a “fair,” “ethical” algorithmic system for targeting elderly individuals for processing into a “high-nutrient slurry” \cite{Keyes2019-dt}. After walking through a detailed and thorough procedure for ensuring a more ethical algorithmic system - including input from salient stakeholders -  the authors suggest that “sometimes the problem is not how the sausage gets made, but that they’re making people into sausage” (p.7) \cite{Keyes2019-dt}.

Green and Viljoen suggest that Andre Gorz’s schema of “reformist” versus “non-reformist reforms” likewise helps to describe this split between affirmative and critical design, or between data ethics and data justice: 
\begin{quote}
While a reformist reform “subordinates its objectives to the criteria of rationality and practicability of a given system and policy,” a non-reformist reform “is conceived not in terms of what is possible within the framework of a given system and administration, but in view of what should be made possible in terms of human needs and demands.” (p.26) \cite{green2020algorithmic} 
\end{quote}
\subsection {Refusal \& Abolition as Generative}
A related thread of literature explores how designers and data professionals might respond when called to build systems that, metaphorically, are making people into sausage. These scholars point towards the feminist, abolitionist practice of “refusal” \cite{Garcia2022-iz}. Broadly, “refusal” looks like the individual and collective tactics that users and designers employ to step outside of the prescribed “problem frame” of fundamentally extractive and unjust systems. As Barabas summarizes, “In its most potent form, refusal operates as a framework for renegotiating the terms of engagement” \cite{Barabas2022-yr}. For data professionals and designers immersed in discourses of responding to these injustices through increased access, inclusion, or transparency \cite{James2023-oj}, “refusal” looks like side-stepping the imperative to improve these existing technical infrastructures and systems \cite{Hoffmann2021-uh}. Users and those directly impacted by harmful data systems can also practice refusal; Zong and Matias powerfully argue that users' practices of refusal should also be understood as a kind of design-practice \cite{Zong2023-ns}. 

The framework of refusal as articulated within critical data studies and feminist HCI literature explicitly draws from abolitionist scholarship, including Ruha Benjamin’s “Race After Technology: Abolitionist Tools for the New Jim Code” (2019), and from abolitionist geographer Ruth Wilson Gilmore. Central to abolitionism is the idea that abolishing prisons and policing is not simply about destroying current carceral systems, but an imaginative and generative practice of building alternative forms of safety and care. Refusal, likewise, is understood as a “strategic and generative act” \cite{Barabas2022-yr}.

\subsection {Non-Reformist Reforms as a Practical Framework for Data Justice}
Seeking to expand our theoretical and practical capacities for “doing” refusal - doing data justice - this paper takes up the framework of “non-reformist reforms,” specifically as developed by abolitionist scholars and activists. Every day, prison and police abolitionists struggle to identify paths forward for addressing the immediate harms of mass incarceration - over 1.9 million people being held in cages in the United States alone \cite{Sawyer2024-na} - and racist and violent policing. Especially when it comes to collaborating with lawmakers or the non-profit industrial complex, it can be particularly difficult to do this harm reduction work in ways that are not reformist: “Unfortunately, many remedies proposed for the all-purpose use of prisons to solve social, political, and economic problems get caught in the logic of the system itself, such that a reform strengthens, rather than loosens, prison’s hold” (p.242) \cite{Gilmore2007-cv}. For instance, immense efforts are being made by organizers in the New York City area to close the Rikers Island jail, which has been the site of sensational abuse, scandal, and violence. Yet the state’s proposed path for closing Rikers involves opening four new jails across the City \cite{Rubinstein2024-zi}: a troubling but extremely common example of the ways that efforts to reduce the harms of prisons and policing often reproduce prisons and policing in new ways.

This problem is likely familiar to anyone who has ever attempted to make change within any kind of institutional system or structure: what incremental changes are possible today that do not reproduce the very same problem we are trying to solve? How do we differentiate between reformist reforms and non-reformist reforms? In what follows, I analyze proposed efforts to automate prior authorization through the lens of non-reformist reforms, providing a concrete example of what it looks like to apply this framework to emerging data- and technology-driven reforms of harmful and extractive systems.

\section{Method}
This case study draws from 18+ months of ethnographic research, including interviews with health informatics professionals and federal and state policy makers; fieldwork at formal and informal events held for health IT professionals and regulators, including national and regional conferences, social events, workshops, and board meetings; and discourse analysis of emerging health data regulations and standards, as well as webinars, white papers, and other industry coverage of these regulations and standards. This study was approved via an expedited IRB review process. Interviewees’ and workshop attendees’ identities have been anonymized and specific organizational affiliations kept confidential.  

Drawing from theoretical frameworks of economic sociology and STS, this fieldwork is focused on exploring emerging health data regulations as a political economic phenomenon. Through situational mapping \cite{Clarke2003-lf}, I identified a number of different case studies that illustrate how emerging technical regulations are closely implicated with a particular mode of managing care, and managing healthcare markets. Prior authorization regulations, which are the focus of this paper, are just one case study within this overall situational map. This case study of prior authorization provides one of many possible entry points for examining the ways that social and economic value(s) are enacted through emerging health data regulations. 

The findings presented in this paper emerge primarily from three different sources. First, I draw from fieldwork at a two-day workshop, hosted in summer 2024 by a national health informatics professionals association following the release of the final prior authorization regulations earlier in the year. This workshop explicitly focused on exploring the impact and implications of these new prior authorization regulations, and included around 200 attendees and presenters from federal regulatory bodies, health insurance organizations, healthcare provider organizations, national advocacy organizations (such as the American Medical Association), and a wide variety of technical and administrative vendors implicated in the prior authorization process. I also did follow-up interviews with several event attendees in the weeks following the event. 

Additionally, the findings outlined here draw from a close analysis of the Center for Medicaid \& Medicare Services (CMS) Interoperability and Prior Authorization Final Rule (CMS-0057-F), released in January 2024, as well as the public comments and industry coverage about this regulation and previous proposed rules (released in December 2022 and December 2020). I also draw from technical documentation for updated data exchange standards that are referenced in these regulations. This approach draws from a Critical Technical Discourse Analysis (CTDA) method \cite{Brock2018-je}, and brings together a close analysis of the regulatory and technical artifacts in and of themselves with an articulation of the meaning and intent as understood by their designers and implementers. 

These fieldnotes, interview transcripts, and discourse analysis of regulatory, technical, and industry coverage documents were collectively annotated and coded to explore the following research questions:
\begin{itemize}
    \item What do different actors within the field understand to be the goal or purpose of prior authorization?
    \item What do different actors within the field  (e.g., policymakers, health IT professionals from insurance and healthcare provider organizations) understand to be the goal or purpose of emerging prior authorization reforms, particularly the reforms focused on modernized data exchange via APIs?
    \item What are the likely impacts of these prior authorization reforms, particularly the forms focused on modernized data exchange via APIs, according to experts within the field?
\end{itemize}

This analysis surfaced the idea of the “bridge to nowhere” as an analytic that helped to describe both the likely limited impact of these prior authorization reforms, as understood by participants themselves, and what I understood as the gap between the stated purpose or intent of prior authorization (e.g., to avoid extractive, harmful, or exploitative forms of care by providers) and its operationalization (e.g., a demonstrably harmful and extractive gatekeeping of care by insurance providers).

\section{Harm Reduction Through Automation, Datafication: A Case Study of Prior Authorization Reforms}

This case study explores emerging federal regulations pertaining to prior authorization: a bureaucratic process whereby insurance companies require review and approval for coverage of certain procedures or treatments. This bureaucratic gatekeeping is intended to mitigate the use of unnecessary, unsafe, and/or unnecessarily costly medical interventions. Yet this process has been demonstrated to cause harm to patients via delayed care, or coverage rejections for needed care - and to create additional administrative hurdles for already overstretched healthcare providers. As a result, prior authorization has been the target of widespread criticism from groups like the American Medical Association, and legislators have called for federal reforms to prior authorization to reduce the burdens of this process for both providers and patients. 

The typical prior authorization workflow involves a doctor recommending a treatment to a patient; the provider organization will then check to make sure that this treatment will actually be covered by the patient's insurance. Some treatments are not covered at all, while others require the provider to first provide documentation to the insurance company about the “medical necessity” of this specific treatment. This documentation is reviewed, variously, by medical practitioners employed by the insurance company - sometimes from the same field of medicine, sometimes not - or by a third-party firm. Medical necessity is often determined by pre-written policies following industry “best practices” - though a common complaint among providers and patients is that these standardized policies do not account for the wide variation in human lives and illnesses, and the medical discretion a provider uses when attempting to treat a real patient, not a “standard” patient. 

Even in the event that coverage is approved, this process requires significant documentation labor by the provider organization, and often introduces weeks-long delays in receiving medical treatment. Sometimes, a claim is outright denied, and a provider and patient must either choose an alternate course of treatment or begin the appeals process, introducing further delays to treatment. Sometimes, the insurer may request additional documentation before they make a decision. Other times, an insurer may require the patient to first try a less-costly or more "traditional" form of treatment and demonstrate that this treatment is unsuccessful, before they can be approved for the treatment suggested by the provider originally. 

Prior authorizations essentially function by introducing "friction" into the utilization of higher-cost forms of care: it is a soft barrier rather than a hard wall. But these frictions can be life-threatening. One recent study of 178 patients with cancer found that 73\% reported a delay of 2 or more weeks, and 22\% did not receive the treatment recommended by their doctor due to prior authorization denials/delays \cite{Chino2023-in}. An interview with radiation oncologists about the impacts of prior authorization highlights the impacts of these delays for cancer patients \cite{Marshall2024-jc}: a 21-year old patient with a rare tumor was recommended for surgery paired with a new form of "proton therapy." During the year-long delay waiting for the new form of treatment to be approved, the tumor came back, and the patient ultimately needed a tracheostomy and a feeding tube. This patient’s doctor is quoted as saying, "Unfortunately for this patient, it went from a potentially curable situation to a likely not curable situation… I wanted to cry every day that she waited'' \cite{Marshall2024-jc}. 

Rapidly-growing cancer is one of the more dramatic examples of the impacts of prior authorization on patients - but the slow train-wreck that many chronically ill patients experience is just as troubling. Often, patients report needing to go through the same battle over treatments multiple times per year, such as when a prescription gets refilled or when they change providers, or when insurance companies tweak their plans. Prior authorization also impacts healthcare providers; a recent survey by the American Medical Association found that physicians and their staff spend an average of 12 hours each week completing prior authorization documentation, and 95\% of physicians responded that prior authorization "somewhat" or "significantly" increases physician burnout \cite{noauthor_2024-oi}. 

Responding to concerns about the harms of prior authorization, the federal government has recently taken steps to reduce the burden of prior authorizations on patients and providers. In March 2024 the Center for Medicare and Medicaid (CMS) released new regulations requiring insurance companies who manage federal plans, such as Medicare Advantage, to modernize the data infrastructure behind their approval processes and provide greater transparency about their decisions to patients, providers, and the federal government, and reducing the time that insurance companies have to make a decision (e.g., from 14 to 7 days, depending on the type of decision). 

These interventions are described under the banner of “burden reduction”: that is, reducing the burden of prior authorization processes for healthcare providers and patients through transparency and improved efficiency in the review process. Importantly, the provisions updating the data infrastructure behind approval processes are explicitly meant to be a step towards enabling the automation of these decision-making processes. Currently, the prior authorization workflow is largely carried out via the exchange of faxes or digital documents between providers and insurers. These new regulations require insurance companies to develop APIs through which the necessary information can be exchanged in a more granular way, and in real-time. Automated information exchange and decision-making via APIs is understood as the "north star" in terms of reducing the burden of prior authorization.

These changes will not fully go into effect until 2027. At this moment, health IT professionals and executives within both payer and provider organizations are assessing how these regulations will impact their workflows and data infrastructures. As I explore further in the following sections, many health IT professionals and health executives charged with implementing these changes are skeptical that these data modernization requirements will have the intended effect of harm reduction and burden reduction for patients and providers, with several describing their work as building a bridge to nowhere. The following case study outlines in more detail the overall design problem that the federal government is intending to solve through datafication and automation and sets the stage for evaluating how a non-reformist reforms framework might help us to avoid building a “bridge to nowhere.” 

This case study is broken into three sections, outlining:
\begin{itemize}
    \item How prior authorization works currently, and its intended purpose in managing "low value care";
    \item How regulators imagine datafication and automation as a tool to reduce the harm/burden of prior authorization processes; and
    \item Why these automation/datafication efforts are understood as a “bridge to nowhere” by some health IT professionals.
\end{itemize}

\subsection{The Goals of Prior Authorization: Insurance Companies as Managers of "Low Value Care"}
Regulators' and insurance companies’ stated rationale for prior authorization is to reduce “unnecessary” or “wasteful care.” The more treatments and interventions a doctor provides, the more money a healthcare organization makes. To mitigate unnecessary treatments and interventions that benefit healthcare providers financially but do not have a big impact on patients’ health outcomes, insurance companies identify certain procedures and treatments that require a doctor to demonstrate “medical necessity” before they agree to cover the treatment. A typical insurance policy outlines which kinds of services or treatments are covered or not, and which require prior authorization and/or a “co-pay” by the patient - another mechanism intended to mitigate “unnecessary” and costly forms of care. Some services or treatments are conditionally covered, if the provider can provide a proof or rationale that it is needed. Prior authorization may be required, for instance, in cases of expensive surgical procedures or for the use of non-generic medications when a generic is available. Additionally, prior authorization requests submitted by a provider do not just yield a yes/no response: often the insurer requires the provider to first try less costly interventions, called “step therapy,” and then demonstrate the failure of those interventions before they can get approved for the more costly intervention. 

Prior authorization is intended as a way to reduce what is described by insurers as “low value care”: that is, costly or unnecessary care with a relatively low “return on investment” in terms of improved patient health.  Broadly, prior authorization reduces costly care in a few different ways. First, through outright rejections, insurance companies effectively require patients to pay for treatments themselves, or to seek other, less costly forms of treatment. Second, by introducing time-consuming prior authorization requirements, insurers are able to deter providers from using more costly interventions, even if those interventions are likely to be approved given proper documentation from the provider. This is known as the “sentinel” effect \cite{Gupta2024-ha}. Finally, by introducing friction into the process, there is a significant amount of attrition around more costly forms of treatment. For instance, patients may not be willing to pursue time-consuming step therapy treatments. Additionally, there is evidence that while only a small fraction of denied prior authorization requests are appealed, a high number of those appeals are ultimately approved \cite{Biniek2024-rs}. This suggests that the cause of these denials is simply to deter the use of costly care interventions, regardless of their “medical necessity.”

The use of prior authorization by both private insurance companies and public programs such as Medicare and Medicaid has expanded significantly since its introduction in the 1980s \textbackslash{}cite\{Institute-of-Medicine-US-Committee-on-Utilization-Management-by-Third-Parties1989-lt\}, and has recently become a site of increased political contestation and scrutiny. Organized medicine, including the American Hospitals Association and the American Medical Association, has pushed back about the time-consuming paperwork and delays to care that current prior authorization workflows entail, with negative impacts on both patients and care providers. In a 2023 survey of 1,000 practicing physicians, the AMA found that 56\% reported that issues in the prior authorization process sometimes led to patients abandoning a recommended course of treatment, while 19\% of physicians reported that prior authorization had led to a patient’s rehospitalization. Responding to these concerns about the impact on patients and providers, states and the federal government have sought to reform prior authorization. The next section outlines the goals and values articulated through these specific reforms.

\subsection{The Goals of Prior Authorization Reform: “Burden Reduction” Through Datafication, Automation}

Within the past few years, prior authorization has been a target of significant reform at both the federal and state levels. Many of these interventions are focused on modernizing the prior authorization workflow in order to increase its efficiency and reduce the administrative burden on patients and providers. In June 2023, 233 representatives and 61 senators signed a letter advocating for CMS to finalize prior authorization reforms that include “a mechanism for real-time decisions”  \cite{United_States_Senators_undersigned2023-dc}. The Senate letter notes that “the health insurance industry has been actively working towards real-time decisions through automation and artificial intelligence from end-to-end” (p.2) - emphasizing that “end-to-end” automation would be central to providing relief for the harms of prior authorization. 

In December 2022, the Center for Medicare and Medicaid Services (CMS) released a proposed rule, “Advancing Interoperability and Improving Prior Authorization Processes,” for public comment and feedback. This proposed rule outlined potential requirements for payers to build APIs enabling the more seamless exchange of information between providers and payers about prior authorization requirements, requests, and documentation. As of 2023, 37\% of the information sent by medical plans for the purposes of prior authorization was relayed manually, via phone, fax, email, or mail \cite{Unknown2024-yw}. (Fig. 1 provides a basic overview of the discrete information transactions involved in the prior authorization workflow.)

\begin{figure*}
    \centering
    \includegraphics[width=.75\linewidth]{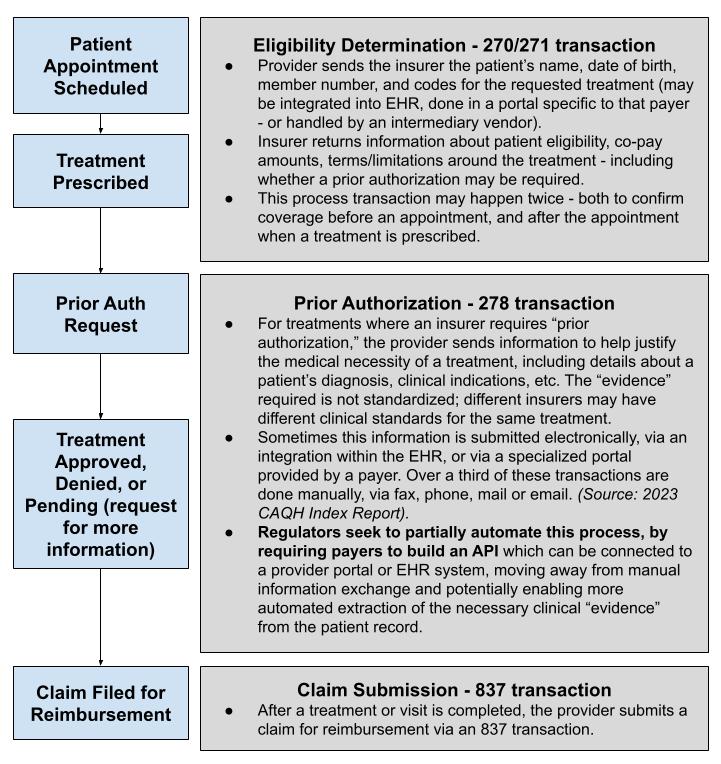}
    \caption{Overview of standardized "X12" information transactions underlying the prior authorization workflow}
    \Description{A flow-chart illustrates three main phases of the prior authorization information workflow. First, small boxes labeled "Patient Appointment Scheduled" and "Treatment Prescribed" are associated with a large box labeled "Eligibility Determination - 270/271 transaction". The description for this transaction is: "Provider sends the insurer the patient’s name, date of birth, member number, and codes for the requested treatment (may be integrated into EHR, done in a portal specific to that payer - or handled by an intermediary vendor). Insurer returns information about patient eligibility, co-pay amounts, terms/limitations around the treatment - including whether a prior authorization may be required. This process transaction may happen twice - both to confirm coverage before an appointment, and after the appointment when a treatment is prescribed." Below this, two boxes labeled "Prior Auth Request" and "Treatment Approved, Denied, or Pending (request for more information)" are associated with a large box labeled "Prior Authorization - 278 transaction". This description for this transaction is: "For treatments where an insurer requires “prior authorization,” the provider sends information to help justify the medical necessity of a treatment, including details about a patient’s diagnosis, clinical indications, etc. The “evidence” required is not standardized; different insurers may have different clinical standards for the same treatment. Sometimes this information is submitted electronically, via an integration within the EHR, or via a specialized portal provided by a payer. Over a third of these transactions are done manually, via fax, phone, mail or email. (Source: 2023 CAQH Index Report). Regulators seek to partially automate this process, by requiring payers to build an API which can be connected to a provider portal or EHR system, moving away from manual information exchange and potentially enabling more automated extraction of the necessary clinical “evidence” from the patient record." Finally, a box labeled "Claim Filed for Reimbursement" is associated with a box labeled "Claim Submission - 837 transaction", which says: "After a treatment or visit is completed, the provider submits a claim for reimbursement via an 837 transaction."
}
    \label{fig: Fig. 2}
\end{figure*}

CMS’ final rule, released in January 2024, also includes new requirements for insurance companies beyond data modernization that are intended to reduce the harm of prior authorization \cite{Centers-for-Medicare-and-Medicare-Services2024-vn}. The final rule also attempts to increase the transparency and accountability for prior authorization practices by requiring insurers to provide providers and patients with a reason for prior authorization denials, and to publicly post summary metrics about their use of prior authorization. These regulations also reduce the time frame allowed for insurers to respond to prior authorization requests, e.g., from 14 to 7 days for non-expedited requests. Overall, these regulations do not challenge insurance companies’ right to determine what is or is not “low value” or “medically necessary” care. CMS sees the prior authorization "problem" as a user experience issue: a matter of “burden reduction.” Through data modernization requirements, they seek to improve the user experience and mitigate the harms of prior authorization to patients and providers through faster, more scalable forms of data exchange and automation.

In summer 2024, a national health IT professionals organization held a two-day workshop dedicated to exploring the impact of emerging prior authorization regulations. Speakers included federal regulators and standards-makers responsible for developing these regulations - as well as executives and data professionals from insurance companies, healthcare provider organizations, and a wide range of intermediary vendors involved in the prior authorization workflow. 

At this workshop, attendees were presented with a success story about the potential impact of improved data infrastructures for prior authorization. A technology executive from a provider organization shared a story about their close collaboration with an insurer in order to implement automatic, API-based prior authorization requests. The health provider executive described how doctors in her organization reacted to the faster, more automated prior authorization process, which allowed them to get nearly real-time responses to their prior authorization requests. She cited statistics about the reduced hours of labor required, described how providers in her organization called this the “easy button” - and cited requests from doctors and nurses to route more of their workflow through this new “easy button.” 

This user experience success story was intended as a motivational example for the other healthcare data executives and professionals in the room about the potential positive impacts of increased automation and real-time data exchange around prior authorization. Yet following this demo, much of the subsequent panel discussion revolved around whether this success story was realistic for other organizations with fewer resources or different business models. Many in the room were skeptical that they could ever replicate the experience of the “easy button.” The next section explores the many critiques and concerns that health data professionals raised about the kind of impact that these data-modernization efforts would have in terms of reducing provider burden.

\subsection{“A Bridge To Nowhere”: Health Data Professionals' Perspectives on Prior Authorization Burden Reduction }

During this two-day workshop for health industry executives and data professionals, many described how complying with these regulations felt like building a “bridge to nowhere.” They cited numerous organizational and technical barriers that would reduce the likely positive impact of these regulations. One major common concern pertained to the limited scope of these regulations, which only apply to a few specific types of federally-funded insurance plans, and which only apply to decision-making around medical services, whereas the overwhelming majority of prior authorization requirements pertain to pharmaceuticals.

Professionals also highlighted the large number of different actors that would have to collaboratively move in sync to achieve an improved user experience, many of whom are outside of the direct purview of these regulations. These intermediaries include the vendors that support prior authorization on both the provider and insurer sides, and the electronic health record companies that would be the primary interface for providers seeking to find information about prior authorization requirements or submit a prior authorization request. Significantly, provider organizations are not required to adopt API-based information and PA requests, and providers said that they were reticent to build out new information infrastructures until they saw clear return-on-investment in terms of significantly reducing the burden of prior authorizations. This, they pointed out, would largely rely on the benevolence of insurance companies in going above and beyond simple compliance to try to more intentionally reduce the burden of prior authorization by responding in real-time, or enabling the use of API interfaces across all their lines of business, not just the federally funded programs. In this regard, both ends of the intended information “bridge” were skeptical about the other side.

A button (\figureautorefname{2}) distributed at this workshop emphasizes the idea that improved data exchange via APIs can help to rectify the tensions and conflicts around prior authorization. The button says "PRIOR AUTH PEACEMAKER," and depicts a cat in a white coat and boxing gloves - presumably a healthcare provider - swinging at a dog in a suit - perhaps representing the insurance company or other intermediary vendor. The dog, smiling, absorbs the cat's blows. This obviously silly button nonetheless gives insight into the mental model of some in the room: that the conflict over prior authorization can be defused through new information infrastructures that make the process more seamless for providers. Regulators construe modernized data exchange via APIs as the technical fix for the conflict over who gets to make decisions about care. 
\begin{figure}
    \centering
    \includegraphics[width=0.50\linewidth]{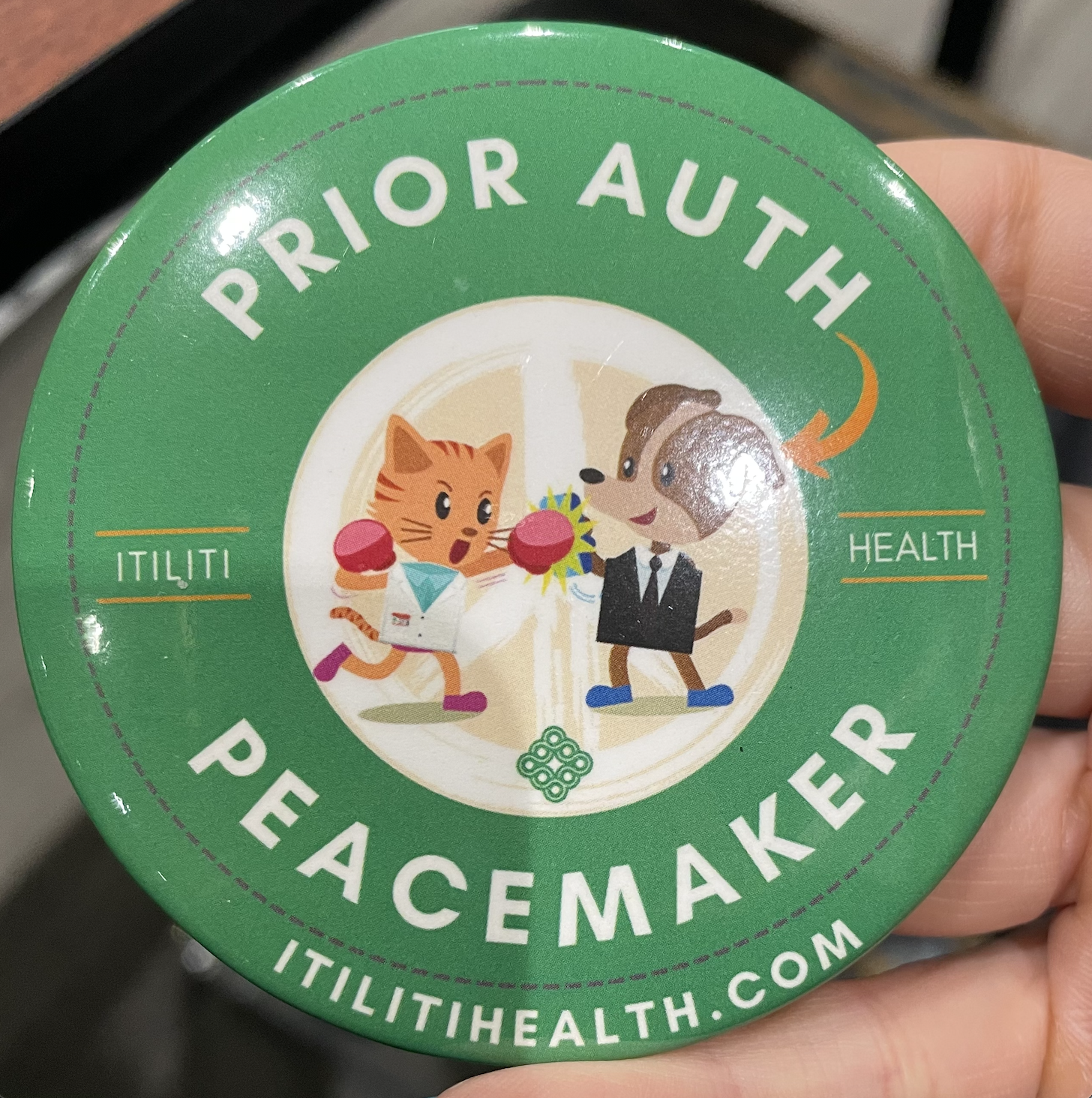}
    \caption{A "Prior Auth Peacemaker" pin distributed at a workshop on emerging prior authorization reforms attended by health data professionals. Photo taken by the author.}
    \label{fig:Fig.1}
\end{figure}

Yet many of the health data professionals within this workshop, from both payer and provider organizations, expressed deep reservations about the ability of these reforms to the data infrastructures underlying prior authorization to make a meaningful impact on the lives of patients or overstretched healthcare providers. These concerns were largely about the organizational and technical complexity the reforms required, and the likelihood of failure in a context where change is both immensely expensive and against the immediate financial interests of the companies implementing the changes. 

Even if we imagine these complex reforms being implemented seamlessly and with full-throated enthusiasm from the necessary players, through the lens of "non-reformist reforms" we might understand these proposed changes as a “bridge to nowhere” in another sense entirely. These reforms ultimately reaffirm insurance companies as an appropriate adjudicator of “medically necessary care.” Rather than directly addressing the underlying conflict around whether insurance companies’ own profitability is the best proxy for ensuring the equitable and responsible distribution of limited healthcare resources, federal regulators seek to act as a “peacemaker” by making insurance companies' control over care decisions less visible and less disruptive through increased automation of the prior authorization workflow. In the next section, I explore how the framework of non-reformist reforms can help us to imagine alternative pathways for reducing the harm of prior authorization, without extending and reproducing the power of insurance companies over care. 

\section{“A Bridge to Nowhere”: Prior Authorization Automation as a Reformist Reform} 
If we set aside the many operational concerns raised by the health technologists in the room that day, we can speculate that perhaps a widespread “easy button” for prior authorization would help achieve the intended goals of harm reduction: reducing the workload for over-worked healthcare providers, and reducing harmful delays in care for patients. However, this easy button solution would also secure, rather than challenge, insurance companies’ power to make unilateral decisions about “low value” or “medically necessary” care. Additionally, by reducing the administrative friction of prior authorization, insurers may be enabled to require prior authorization for a broader set of treatments and services. In this regard, prior authorization policies that seek to improve the user experience represent a retrenchment and potential expansion of insurance companies’ unilateral power to define “low value care.”

Intervening to improve the user experience of prior authorization takes a fundamentally conservative approach to existing power relations. In their messaging about these regulations, policymakers reaffirm that “there is a place for prior authorization” in terms of governing care. The problem-frame, and its solution, is specifically focused on the last mile: fixing the user experience of an existing decision-making system. Important questions remain unasked: is prior authorization an effective design mechanism for reducing “low value” care? By what definition of low-value care? For-profit insurance companies may well have a different definition of “low value care” relative to the patients and tax-payers who both pay for and receive that care, yet the state treats insurers’ definitions of low value care as an appropriate proxy for the nation’s. 

This mechanism for utilization management is not applied to the other major public insurance programs Medicare and Medicaid. Likewise, there is not consistency in the treatments that require prior authorization across different insurers \cite{Gupta2024-ha}. This suggests that there is not strong consensus or evidence about the effectiveness of this mechanism for managing exploitative or costly forms of care, even across different insurance companies. It makes little sense to build data infrastructures to automate decisions about what kind of care is valuable, given this lack of consensus even among the insurance companies - let alone between the state, patients, providers, etc. I argue that these data governance interventions represent a “bridge to nowhere,” not only in terms of technical and organizational barriers described by professionals in the field -  but in terms of developing an infrastructure that enables the redistribution of power over decisions about the best way to allocate healthcare resources to patients, rather than to insurance companies interested in maximizing shareholder value. 

In the next section, I reflect on the broader implications of this case study for policymakers and designers interested in advancing data justice. I analyze the focus on prior authorization user experience as an example of a “reformist reform,” and reflect on the possibility of interventions that would represent “non-reformist reforms,” drawing here from terminology advanced by abolitionist and legal scholars. This framework points towards the possibility of designing interventions that both alleviate immediate harms while also building a bridge towards the dismantling of unjust systems.

\section{Discussion} 
Having provided an overview of prior authorization, its goals and proposed reforms, the following section explores how we might view these reforms differently through the lens of non-reformist reforms. I then propose four tenets of non-reformist design more broadly. 

\subsection{How Not to Build A Bridge to Nowhere: Refusing Prior Authorization}

Ultimately, the introduction of APIs and modernized data infrastructure in order to move towards automated prior authorization is a push that increases, rather than decreases, the power and control that insurance companies wield over healthcare decisions. While it would potentially improve some patients’ and providers’ experience with the prior authorization workflow, if successful it may in fact enable insurance companies to further expand the domain of treatments that require prior authorization, further extending insurance companies’ domain of authority to determine "low value" or "medically necessary" care.

Abolitionist organization “Critical Resistance” has a rubric to help activists evaluate the difference between reformist and non-reformist reforms \cite{Critical-Resistance2020-bo}. This rubric asks questions like: “Does this reduce the number of people imprisoned?”, “Does this reduce the reach of jails, prisons, and surveillance in our everyday lives?”, “Does this create resources and infrastructures that are…accessible without police and prison guard contact?” In the domain of prior authorization, a similar set of questions might look like: “Does this reduce the control that for-profit entities have over the care that a patient receives?” - or, “Does this help to ensure that health resources are distributed equitably, and in a way that is collectively decided by people and communities, not corporations?”

Using data infrastructures to improve the efficiency of prior authorization does not meet any of these criteria. However, it is important to highlight that data modernization interventions are only one piece of the prior authorization reforms that have recently been introduced. Other aspects of recent regulations - including limiting the scope of treatments that can be subject to prior authorization - look more like a non-reformist reform, to the extent that they reduce the power and control of for-profit entities and corporations over the distribution of care. Other dimensions of these regulations exist in a kind of gray area, mitigating the harms of prior authorization in ways that do not obviously expand the power or control of insurance companies over care decisions - for instance, by reducing the amount of time that an insurance company has to respond to a prior authorization claim. We might understand these interventions as aligning with Creary’s definition of “bounded justice” \cite{Creary2021-fh, El-Azab2023-vz}: interventions that seek to advance health justice in a context fundamentally structured around social inequity.

Regulators’ attempts to mitigate the harms of prior authorization through improved data infrastructures do not call into question the basic premise that insurance companies have a role to play in determining whether proposed treatments are “low value” or “medically necessary.” Within the U.S. healthcare system, health insurance companies have been designated as the “check and balance” on overtly extractive or exploitative tendencies of healthcare provider organizations. Yet there is relatively little by way of checks and balances on the extractive or exploitative tendencies of the insurance companies themselves. A more radical kind of “non-reformist reform” would look like abolishing prior authorization entirely, based on the understanding that insurance companies’ own bottom line is not a sufficient proxy for determining the most equitable distribution of healthcare resources.

In my fieldwork and interviews, even the most vocal critics of prior authorization were reluctant to suggest that there is no place for prior authorization within our healthcare system. And yet - the widespread push for prior authorization reforms, including among powerful actors like the American Medical Association, suggests that there is momentum that could be channeled into non-reformist avenues of change. The recent outpouring of public rage in the wake of the targeted assassination of UnitedHealthCare’s CEO highlights that there may yet be more broad political support for measures that reduce insurance companies’ scope of control over care. Regulations that focus on improving the “user experience” of prior authorization seek to quell, rather than channel, this broad discontent among patients and healthcare workers.

In the remaining sections of this paper, I move beyond this specific case study to explore how the framework of “non-reformist reforms” helps us to imagine alternative ways to mitigate harm and redistribute power.

\subsection{Designing Data Justice: Four Tenets for Designing Non-Reformist Reforms}

Inspired by abolitionist advocacy organization Critical Resistance's rubric for reformist versus non-reformist reforms, I have outlined below four tenets for non-reformist design, broadly speaking. In the following sections, I briefly reflect on the ways that these tenets echo and amplify themes from existing literature on design justice, abolition, and refusal in the context of critical data studies and feminist HCI.

\begin{itemize}
    \item \textbf{Expanding the problem frame}
    \begin{itemize}
        \item Is the problem being located within individuals or technical systems - or within structural power differentials or inequities?
        \item Follow the problem “upstream”: are we solving a problem that shouldn’t exist?
    \end{itemize}
    \item \textbf{Expanding the methods or strategies for intervention:}
    \begin{itemize}
        \item Is the proposed intervention working narrowly within the confines of interventions or methods that are currently trendy or fund-able? Or are a broader set of tools and techniques considered?
    \end{itemize}
    \item \textbf{Expanding the power of users, marginalized groups}
    \begin{itemize}
        \item Does this intervention increase the involvement of those who are most impacted or marginalized in defining the social good to be achieved and the best mechanism for achieving it?
    \end{itemize}
    \item \textbf{Reducing the power of already-powerful actors}
    \begin{itemize}
        \item Does this reduce or lessen the role of already-powerful organizations in defining the social good to be achieved and the best mechanism for achieving it?
        \item Does this reduce the scope for already-powerful actors to structure, extract, and control data flows?
    \end{itemize}
\end{itemize}

\subsubsection{Expanding The Problem-Frame}
Most of the regulators and health IT professionals working on implementing these changes were prepared to evaluate emerging prior authorization reforms within the problem-frame of “improving the user experience.” Relatively few questioned the larger goals or values of prior authorization itself, or even whether there was clear evidence of the necessity of prior authorization as a tool for achieving those goals (e.g., reducing "low value care"). 

Passi and Barocas, in their ethnographic research with a corporate data science team, highlight problem-formulation as a site of extensive normative decision making, often unexamined by data scientists themselves \cite{Passi2019-fy}. Similarly, within the site of prior authorization reform, the problem-formulation stayed within the scope of “user experience” around prior authorization. Designing non-reformist reforms, in contrast, requires a careful investigation of the terms of the “design challenge” itself. Just as focusing on “how can we reduce racial biases in policing?” is a problem-frame that takes as a given that police are a good way to ensure safety in our communities, focusing on improving the user experience of prior authorization is a problem frame that takes as a given that limited care resources are being unnecessarily wasted by providers and patients, and that for-profit insurance companies are a good arbiter of that “waste.” 

Within HCI scholarship, Dencik, Hintz and Cable likewise highlight the central importance of redefining the problem frame to the work of data justice. Speaking with social justice activists in the wake of the Snowden leaks, the authors show how privacy and surveillance concerns among the activist community are treated as a discrete techno-legal problem, to be addressed by a subset of experts within the group - rather than considering the implications of data-driven governance more broadly for the possibility of social justice work and activism. They propose the term “data justice” to describe this broader conceptualization of the structural and systemic concerns of data-driven governance \cite{Dencik2016-vr}. 

In a related vein, some of the HCI and critical data studies scholarship introduced earlier highlights the issues that arise when a problem is framed specifically as a \textit{technical} or computational problem. D’Ignacio and Klein, for instance, point toward an inherent conservatism in data ethics approaches that narrowly locate “the source of the problem in individuals or technical systems.” Keyes et. al’s article similarly highlights how simple checklists for fair, ethical, and transparent algorithms are perfectly designed to produce an apparently “ethical” technical process that in no way addresses the fundamentally violent purpose of the underlying system or infrastructure. This brings us to the second tenet of non-reformist design.

\subsubsection{Expanding the Methods or Strategies for Intervention}
In addition to expanding the problem frame, designing for data justice requires broadening the scope of methods, strategies, and tactics for intervention and problem solving. Legal scholar Amna Akbar, in exploring the possibility for legal professionals to contribute to the development of non-reformist reforms, examines how her own discipline, law, has historically contributed to processes that concentrate power and allocate decision-making to experts - even in cases where they are pursuing a social good  \cite{Akbar2022-ko}. Akbar highlights the critical importance of “horizons beyond legalism.” Green and Viljoen draw a parallel between the limits of legalism for achieving non-reformist reforms and the limits of computational approaches \cite{green2020algorithmic}, which tend to re-frame the fluid, complex social problems of the world to make them amenable towards universal, "objective," scalable solutions. 

In a similar vein, Ruth Wilson Gilmore explicitly points towards the role of professionalization and funding streams in perpetuating the production of reformist-reforms that adhere to an overly narrow set of interventions which are not fit to make systemic or structural change. She argues that professionalization "has made many committed people so specialized and entrapped by funding streams that they have become effectively deskilled when it comes to thinking and doing what matters most” (p.242) \cite{Gilmore2007-cv}. What might it look like for HCI scholars to move beyond the constraints of professionalization - to put down the hammer for a moment to engage social problems more fully, rather than remaking them into nails?

In “HCI Tactics for Policy from Below,” Whitney et. al provide concrete examples of what this might look like. The authors use as a case study their involvement with community-based organizing and advocacy around smart city interventions, and specifically highlight strategies they employed to “attune HCI to the needs and practices of those excluded from power over widespread technology infrastructures” (p.1) \cite{Whitney2021-vd}. They highlight their work interpreting legal and technical blackboxes, in reading documents and in translating information learned from a hackathon into a policy report. They argue that “solidarity means going beyond the questions that we, our field, or our employers find interesting. It means taking an interest in what others want or need to know” (p.12), and point towards the necessity of engaging in “forms of political struggle that exceed the limits of project duration, grant funding, and other bureaucratic hurdles common in both industry and academia” (p.13) \cite{Whitney2021-vd}.

Ruha Benjamin argues that the term "design justice" re-defines justice in terms of that which is institutionally recognizable or rewarded: "by adding "design" to our vision of social change we rebrand it, upgrading the social change from 'mere' liberation to something out of the box, disrupting the status quo" (p.179) \cite{Benjamin2019-ef}. She suggests that "maybe what we must demand is not liberatory designs but just plain old liberation" (p.179) \cite{Benjamin2019-ef}. Likewise, perhaps a practice of "non-reformist design" should take us beyond what is recognizable as "design." 

In the case of prior authorization automation, we see a complex social problem - how do we make decisions about how to allocate care resources? - flattened and constrained, boiled down into a problem that can be solved through the mechanisms of law and computation. Addressing this complex social problem starts with forming alliances with others who seek to solve this problem, and finding ways to support their work. For example, it might look like partnering with patient advocacy organizations proposing legislation to eliminate the use of prior authorization. It might look like interviewing patients and providers about the impacts of prior auth workflows - including the impact of "faster" workflows that still end up restricting access to care. It might look like sharing these testimonials with legislators and on social media to catalyze a broader recognition of prior authorization as a fundamentally harmful practice that should be abolished rather than reformed. Of course, this is just an example: the most important "method" of non-reformist design is to become an accomplice to the users and groups harmed by the current system.

\subsubsection{Expanding the Power of Users, Marginalized Groups} 
Whitney et. al’s piece also points towards another crucial aspect of data justice practice: to support, through their methods, the community members who are typically excluded, whether structurally or due to the opacity of technical and legal structures, from decision-making about the data infrastructures that shape their lives. Their interventions as HCI scholars are oriented not towards fixing a problem for others, but towards providing tools and information that support their more direct involvement and power over those structures \cite{Whitney2021-vd}. 

Often, in HCI, we identify user discontent as a problem to be solved through a user experience intervention. But user discontent is also an important signal that perhaps a systemic change is needed, and that there are users, patients, and community members prepared to take action for that change, if they are given the tools and structural power to do so. In this sense, to seek data justice may not be a matter of solving the problem at hand, but of enabling users to have the power to solve the problem themselves. 

Akbar suggests that “Non-reformist reforms…embrace antagonism and conflict rather than depoliticization and neutrality” \cite{Akbar2022-ko}. In the case of prior authorization regulations, we see widespread discontent among providers and patients about the harms of insurance companies’ intervention into decision-making about “medically necessary care.” This seems to present a valuable entry point for dialogue and organizing around the structural power that insurance companies wield in determining the optimal distribution of care resources, and for examining how patients can have more power over these decisions. 

Expanding the power of users to make their own decisions is an approach that takes seriously the maxim, “nothing about us without us” \cite{Harbord2021-zh, Spiel2020-vy} - not in the sense of "inclusion" or "participation," which may often involve marginal or performative roles for users - but in terms of changing the structures of power. In the case of prior authorization governance, for instance, perhaps the important design challenge is to ask how people could be more involved in defining the social good and the best mechanisms for achieving it - that is, how do we ensure that there are enough care resources, equitably distributed? How do we minimize treatments given for the sake of profit rather than care?

\subsubsection{Reducing the Power of Already-Powerful Actors} 
Hand-in-hand with the need to expand the power of marginalized users, data justice requires actively curtailing the power of already-powerful actors. In particular, although data can be useful in projects of counter-power and resource allocation, data is an important form of market and social power \cite{Pistor2020-bd}, and it is difficult to avoid creating data infrastructures that do not in some way rely on the expansion of the data-power of already-powerful actors. 

For instance, the prior authorization interventions examined here are intended to reduce harm and burden through automated data transfer between electronic health records and the payer. Ultimately, this is a governance strategy that re-affirms insurance companies’ power to make decisions about care, and furthermore potentially expands the speed and scale at which clinical health record data is shared with insurers. Looking more broadly within the healthcare industry, we already see that health data is significantly structured and controlled by insurance company payment regimes, which dictate the input of diagnostic and treatment codes that will yield the highest rate of reimbursement. This highlights the critical importance of identifying interventions that shift away from, rather than reproduce, insurance companies' power to structure what data is collected, how, and what it is ultimately used for. 

As explored in the previous section, these regulations are designed to enable automated decision making around prior authorization, re-inscribing a system in which decision-making is deferred to “the market” (e.g., insurance companies are able to define “low value care”), to medical experts hired by insurance companies to write clinical policies, and to computational/algorithmic enactments of those expert decision-making structures. What interventions would instead curtail or limit the decision-making by these already-powerful actors, and redistribute this power to patients and their healthcare providers? 

Overall, these four techniques (expand the problem frame, expand the methods or strategies, expand the power of marginalized users, reduce the power of already powerful actors) - which build and draw upon the insights of abolitionist organizing, as well as from the insights of feminist HCI, critical data studies, and critical design - may be helpful as we seek to develop interventions advancing data justice. By applying this framework to the domain of prior authorization regulations design to reduce provider and patient burden, we can see how this intervention at the level of user experience represents a “bridge to nowhere,” working with a constrained set of tools/strategies within a constrained problem-space, seeking to build solutions for users rather than to build user power. This framework helps to counteract the inherent conservatism of interventions at the level of “user experience,” particularly within data infrastructures in domains like healthcare, housing, education, or safety - critical sites for ensuring that our data infrastructure designs do not reproduce a politics of extraction and injustice.

\section{Conclusion} 
This paper explores a particular case study of health data reform, analyzing federal interventions to automate prior authorization decisions by health insurers as a “bridge to nowhere.” These regulations are described by health IT professionals in this field as a bridge to nowhere inasmuch as these interventions are partial, fragmented, and rely on the benevolence and goodwill of insurers in their approach to compliance. Applying the framework of “non-reformist reforms” developed by prison abolition scholars and activists, I analyze these interventions as a “bridge to nowhere” inasmuch as they reflect an intervention at the level of “user experience”: a reformist reform that seeks to mitigate the burden of prior authorization for patients and providers while reproducing and potentially expanding insurance companies’ unilateral control over the definition of “medically necessary care.”

Hoping to identify a horizon "beyond user experience" (as abolitionist scholars have sought a horizon "beyond legalism"), I outline four tenets for non-reformist design and data justice: expand the problem frame, expand the methods or strategies, expand the power of marginalized users, and reduce the power of already powerful actors. I weave together insights from this particular case study, from abolitionist thinking, and from previous scholarship in the domain of feminist HCI, critical data studies, and critical design scholarship to demonstrate the importance and practical usefulness of these four tenets as part of a design and data justice approach. These tenets are crucial for moving beyond the inherent conservatism of data justice interventions that operate purely at the level of “user experience,” and are particularly key for ensuring that our data infrastructures in domains like healthcare, public safety, or education are aligned with the pursuit of data justice. 

\bibliographystyle{ACM-Reference-Format.bst}
\bibliography{sample-base}

\end{document}